\documentclass[12pt]{article}
\usepackage{amsfonts}
\usepackage{amssymb}
\usepackage{epsfig}
\catcode`\@=11
\newcount\hour
\newcount\minute
\newtoks\amorpm \hour=\time\divide\hour by 60\minute
=\time{\multiply\hour by 60 \global\advance\minute by-\hour}
\edef\standardtime{{\ifnum\hour<12 \global\amorpm={am}%
        \else\global\amorpm={pm}\advance\hour by-12 \fi
        \ifnum\hour=0 \hour=12 \fi
        \number\hour:\ifnum\minute<10
        0\fi\number\minute\the\amorpm}}
\edef\militarytime{\number\hour:\ifnum\minute<10
0\fi\number\minute}
\def\draftlabel#1{{\@bsphack\if@filesw {\let\thepage\relax
   \xdef\@gtempa{\write\@auxout{\string
      \newlabel{#1}{{\@currentlabel}{\thepage}}}}}\@gtempa
   \if@nobreak \ifvmode\nobreak\fi\fi\fi\@esphack}
        \gdef\@eqnlabel{#1}}
\def\@eqnlabel{}
\def\@vacuum{}
\def\marginnote#1{}
\def\draftmarginnote#1{\marginpar{\raggedright\scriptsize\tt#1}}
\def\draft{
        \pagestyle{plain}
        \overfullrule=2pt
        \oddsidemargin -.1truein
        \def\@oddhead{\sl \phantom{\today\quad\militarytime} \hfil
        \smash{\Large\sl DRAFT} \hfil \today\quad\militarytime}
        \let\@evenhead\@oddhead
        \let\label=\draftlabel
        \let\marginnote=\draftmarginnote
        \def\ps@empty{\let\@mkboth\@gobbletwo
        \def\@oddfoot{\hfil \smash{\Large\sl DRAFT} \hfil}
        \let\@evenfoot\@oddhead}
        \def\@eqnnum{(\theequation)\rlap{\kern\marginparsep\tt\@eqnlabel}%
        \global\let\@eqnlabel\@vacuum}  }

\renewcommand{\theequation}{\thesection.\arabic{equation}}
\renewcommand{\thefootnote}{\fnsymbol{footnote}}
\newcommand{\newsection}{    
\setcounter{equation}{0}\section}
\def\appendix#1{\addtocounter{section}{1}\setcounter{equation}{0}
\renewcommand{\thesection}{\Alph{section}}
\section*{Appendix \thesection\protect\indent \parbox[t]{11.15cm}{#1}}
\addcontentsline{toc}{section}{Appendix \thesection\ \ \ #1}}

\def \bi{\bibitem}

\def \ov {\over}

\def \b {\beta}

\def\st {\bowtie}
\def\be{\begin{equation}}
\def\ee{\end{equation}}

\def \st {\ltimes}
\def\sto {\oplus_s}

\catcode`\@=11

\def\bea{\begin{eqnarray}}
\def\eea{\end{eqnarray}}
\def\beann{\begin{eqnarray*}}
\def\eeann{\end{eqnarray*}}
\def\beq{\begin{equation}}
\def\eeq{\end{equation}}
\def\ba{\begin{array}}
\def\ea{\end{array}}
\def\ben{\begin{enumerate}}
\def\een{\end{enumerate}}

 \def\be{\begin{equation}}
\def\ee{\end{equation}}

\font\mybb=msbm10 at 11pt

\def\bb#1{\hbox{\mybb#1}}

\def\bZ {\bb{Z}}
\def\bR {\bb{R}}

\def\bC {\bb{C}}

\def \ov {\over}

\def \ee {\epsilon}

\def \bi{\bibitem}
\def\a{\alpha }

\def \G {\Gamma}

\def \b {\beta}

\def\be{\begin{equation}}
\def\ee{\end{equation}}

\def \bi {\bibitem}

\def \nn {\nonumber}

\begin{document}
\date{November 2002}
\begin{titlepage}
\begin{center}
\vspace{2.0cm} {\Large \bf
 The holonomy of the supercovariant connection and Killing spinors}\\[.2cm]

\vspace{1.5cm}
 {\large  G. Papadopoulos
 and D. Tsimpis}

 \vspace{0.5cm}

 Department of Mathematics\\
 King's College London\\
 Strand\\
 London WC2R 2LS
\end{center}

\vskip 1.5 cm
\begin{abstract}

\noindent 
We show that the holonomy of the supercovariant connection for
M-theory backgrounds with $N$ Killing spinors reduces to a
subgroup of $SL(32-N,\bR)\st (\oplus^N \bR^{32-N})$. We use this
to  give the necessary and sufficient conditions
 for a background to admit $N$ Killing spinors.
We show that there is no topological obstruction for the existence
of up to 22 Killing spinors in eleven-dimensional spacetime.
We investigate  the  symmetry superalgebras of
supersymmetric backgrounds and find  that their structure constants are
  determined by an antisymmetric matrix.
The Lie subalgebra of bosonic generators is related to a real
form of a symplectic group.
We show that there is a  one-one correspondence between
certain bases of the Cartan subalgebra of $sl(32, \bR)$
and supersymmetric planar probe M-brane configurations.
A supersymmetric  probe configuration can involve
up to 31 linearly independent planar branes and
 preserves one supersymmetry. The space of
supersymmetric planar probe
 M-brane configurations  is preserved by an
 $SO(32,\bR)$ subgroup of $SL(32, \bR)$.

\end{abstract}
\end{titlepage}
\newpage
\setcounter{page}{1}
\renewcommand{\thefootnote}{\arabic{footnote}}
\setcounter{footnote}{0}

\setcounter{section}{0}
\setcounter{subsection}{0}
\newsection{Introduction}

The recent classification of 
maximally  supersymmetric solutions in ten- and eleven-
dimensional supergravity theories \cite{gpjfof}
  has raised the hope that the same can be achieved for
 solutions preserving less supersymmetry.
 The classification   was based on  the properties
 of the supercovariant connection, ${\cal D}$,  for maximally supersymmetric
 backgrounds and the use of   Pl\"ucker relations.
 In particular, the property that characterizes the maximally
 supersymmetric backgrounds is that the holonomy of the supercovariant
 connection is equal to
one, ${\rm hol}({\cal D})=1$. This implies that the
 curvature of the supercovariant connection vanishes, ${\cal R}=0$. The vanishing
 of the supercovariant curvature
 together with the field equations is the {\sl full set}
 of conditions that a background has to satisfy in order to preserve
maximal supersymmetry. These conditions were solved with the use of
Pl\"ucker relations and the maximal
supersymmetric backgrounds were determined up to  local isometries.
To summarize, in \cite{gpjfof}
\begin{itemize}
 \item The conditions for maximal
 supersymmetry were derived, and
\item All solutions to these conditions were determined 
up to local isometries.
 \end{itemize}

Many supersymmetric solutions of eleven-dimensional supergravity
are known and have found applications in string and M-theory.
Despite this,
very little has been accomplished towards 
a systematic investigation of supergravity
solutions with less than maximal supersymmetry.
There are 
many reasons for this. To mention some, the Riemann decomposition
theorem does not apply for Lorentzian manifolds, there is no
  Berger-type classification of holonomy groups,
and there is no apparent geometric interpretation
of the supercovariant connection.
Apart from the classification of maximal
supersymmetric backgrounds mentioned above and some other partial results
\cite{ohta}, some progress
towards understanding the backgrounds of
eleven-dimensional supergravity
with one Killing spinor has been made in \cite{pakis}.
In particular, the local form of the
background which
 solves the Killing spinor
equations, for one Killing spinor associated with a time-like
Killing vector, has been constructed.

In this paper, we  describe some
general properties of supersymmetric M-theory backgrounds. In particular,
we  carry out the first part of the programme mentioned
above for maximally
supersymmetric solutions \cite{gpjfof},  to M-theory backgrounds
with $N<32$ supersymmetries. 
We  present a  proof that the (reduced) holonomy of the supercovariant
connection ${\cal D}$ for backgrounds with $N$ Killing spinors reduces to
the semi-direct product of $SL(32-N,\bR)$ with N-copies of
$\bR^{32-N}$,
\be
SL(32-N,\bR)\st (\oplus^N \bR^{32-N})~,
\ee
where $SL(32-N,\bR)$ acts on each copy of $\bR^{32-N}$ in the 
fundamental representation. This result has also been pointed out in \cite{hull}
and in particular that for $N=0$ the holonomy group is $SL(32, \bR)$.
  Therefore a simply connected
M-theory background admits $N$ Killing spinors iff the supercovariant
curvature takes values in the Lie algebra of $sl(32-N,\bR)\sto
(\oplus^N \bR^{32-N})$, where $\sto$ stands for the semi-direct sum.
An eleven-dimensional background admits
precisely $N$ Killing spinors\footnote{We thank M. Duff and J. Liu
for a discussion on this point.} iff 
\be
SL(31-N,\bR)\st (\oplus^{N+1}
\bR^{31-N})\nsupseteq {\rm hol}({\cal D})\subseteq SL(32-N,\bR)\st
(\oplus^N \bR^{32-N})~, 
\ee
i.e. the holonomy of ${\cal D}$ is {\sl contained} in $SL(32-N,\bR)\st
(\oplus^N \bR^{32-N})$ but it is {\sl not contained} in  $SL(31-N,\bR)\st (\oplus^{N+1}
\bR^{31-N})$. 
We also present the {\sl explicit
conditions} that the curvature ${\cal R}$ of the supercovariant
connection should satisfy in order for spacetime to admit $N$
Killing spinors. These together with the field equations
of eleven-dimensional supergravity is the {\sl full set} of
conditions for a background to preserve $N$ supersymmetries.
The proof of the above results  relies on the
properties of the Clifford algebra in eleven-dimensions and, in particular,
to its relation to the group $GL(32, \bR)$.
The groups $SL(32, \bR)$ and $GL(32, \bR)$ have recently been considered
 in the context of symmetries of  M-theory \cite{west, hull} and in \cite{duff}.
In \cite{duff, hull} the holonomy of the supercovariant
connection was also considered for spacetimes which satisfy
several other conditions in addition to those of the existence
of Killing spinors.

We also explore the topological properties of backgrounds with Killing spinors.
We find that that  there is no
topological obstruction for a background
to admit $N< 22$ Killing spinors. The first obstruction
can occur for $N=22$ provided that the top cohomology group
of spacetime does not vanish. The obstruction is identified
as the Euler class of a vector bundle. For a
spacetime with topology $\bR\times \Sigma$, an obstruction can occur at $N=23$.

It was shown in \cite{pakis} that the Killing spinor
equations
\be
{\cal D}\epsilon_i=0~,~~~~~i=1,\dots, N
\ee
 imply certain first order equations for the forms associated
with the Killing spinors. Here we shall show that
these first order equations are the conditions
that the bispinors $\epsilon_i\otimes \epsilon_j$
are parallel with respect to supercovariant connection ${\cal D}$
\be
{\cal D} (\epsilon_i\otimes \epsilon_j)=0~,~~~~~i,j=1,\dots, N~.
\label{rush}
\ee
In particular, we shall show the converse, i.e.~ that the Killing spinor
equations are equivalent to (\ref{rush}) for $i=1$ and $j=1, \dots, N$.

We investigate
some properties of the symmetry superalgebra of backgrounds with
$N$ Killing spinors. We show that the spinorial
Lie derivative commutes with the
supercovariant derivative and use this to prove closure.
In addition, we show that the structure constants of the
commutator
\be
[T_{ij}, Q_k]=f_{ij,k}{}^l Q_l
\ee
can be expressed as
\be
f_{ij,k}{}^p=h_{ki}\delta_j^p+h_{kj}\delta_i^p~,
\label{wittgensteina}
\ee
where $h$ is an 
anti-symmetric (constant) matrix, $Q$ are the odd generators and $T$ are
 the even generators associated
 with the Killing vectors constructed from the Killing spinors.  We also find
  that some of the structure constants of the commutators
of two even generators are expressed
in terms of the structure constants of the commutators of the
even with odd generators. In particular, if the generators $T_{ij}$
 are all
 linearly independent and $h$ is non-degenerate, the Lie algebra spanned
 by the $T_{ij}$s is a real form of a certain symplectic group.

We  further explore the applications of the Clifford algebra
in M-theory to classify all supersymmetric planar-probe M-brane
configurations in $\bR^{10,1}$. The latter are in one-one correspondence
with certain bases of the Cartan subalgebra of $sl(32, \bR)$ spanned
by hermitian traceless matrices.
We show
that the maximal number of {\sl linearly independent} M-branes
that can appear in a supersymmetric configuration is 31. 
This is the same as the
dimension of the Cartan subalgebra of $sl(32, \bR)$.
Such a configuration preserves one supersymmetry.
The proof of the above statements relies on the compatibility
of brane projectors.
We find that $SL(32, \bR)$ does not preserve the space of
supersymmetric planar-probe M-brane
configurations. This space is preserved by an 
$SO(32,\bR)$ subgroup\footnote{
It is  of interest to find applications of this fact in the context
of symmetries of M-theory.}
of $SL(32, \bR)$.

This paper is
organized as follows: In section two, we summarize the properties
of the Clifford algebra in eleven-dimensions and
 we determine the
 holonomy of the
supercovariant connection
for backgrounds  with
$N$ Killing spinors. We also find the conditions that the supercovariant curvature
must satisfy in order for a background  to admit $N$
Killing spinors. In section three, we show that there may be topological
 obstructions  for the existence of $N\geq 22$ Killing spinors and 
we identify
the obstruction for $N=22$.
In section four, we investigate the symmetry superalgebras of supersymmetric
backgrounds. In section five, we classify the supersymmetric planar-probe M-brane
configurations. In appendix A,
we summarize some of the properties of $sl(32, \bR)$
and its relation to the Clifford algebra in eleven-dimensions. In appendix
B, we define spinorial Lie derivatives along vector p-forms and discuss
their applications to supersymmetric backgrounds.

\newsection{Backgrounds with $N$ Killing spinors}

\subsection{The Clifford Algebra in Eleven-Dimensions}

The Clifford algebra\footnote{All these are well-known results and
are summarised e.g. in \cite{harvey}.}
 ${\rm Cliff}(\bR^{10,1})$ as a vector space is
isomorphic to $\Lambda^*(\bR^{10,1})$, ${\rm Cliff}(\bR^{10,1})=\Lambda^*(\bR^{10,1})$,
and so it has dimension $2^{11}$. The Clifford algebra as an algebra is  isomorphic to
\be
{\rm Cliff}(\bR^{10,1})=M_{32}(\bR)\oplus  M_{32}(\bR)~,
\label{cliff}
\ee
where $M_n(\bR)$ is the space of $n\times n$ matrices with real entries.
The two terms in the direct sum are distinguished by the action of
the element $\lambda$ of the Clifford algebra associated with the
volume of $\bR^{10,1}$. This acts as $\lambda=(1,-1)$ on the two factors.
The Clifford algebra ${\rm Cliff}(\bR^{10,1})$ can be written as
$$
{\rm Cliff}(\bR^{10,1})={\rm Cliff}^{{\rm even}}(\bR^{10,1})
\oplus {\rm Cliff}^{{\rm odd}}(\bR^{10,1})~,
$$
corresponding to the decomposition of $\Lambda^*(\bR^{10,1})$ in terms of
even- and odd-degree forms.
It is known that
\be
{\rm Cliff}^{{\rm even}}(\bR^{10,1})=M_{32}(\bR)
\label{ecliff}
\ee
and so $Spin(10,1)\subset {\rm Cliff}^{{\rm even}}(\bR^{10,1})=M_{32}(\bR)$.
The Clifford algebra ${\rm Cliff}(\bR^{10,1})$
has two irreducible (pinor) representations,
both of dimension 32. These
are given by the standard action of $M_{32}(\bR)$ on
$\bR^{32}$,  for each factor in the decomposition (\ref{cliff}).
The even part of the Clifford algebra has a unique irreducible representation
given by the standard action of $M_{32}(\bR)$ on $\bR^{32}$, as it can be seen
from (\ref{ecliff}). This restricts to an irreducible real-spinor (Majorana)
representation $\Delta$
on $Spin(10,1)$. The spinor representation $\Delta$ is equipped with
a (real) non-degenerate skew-symmetric $Spin(10,1)$-invariant inner product $C$, the charge conjugation
matrix\footnote{The inner product $C$ can be used
to identify $\Delta$ and its dual.}. Therefore we have
\be
Spin(10,1)\subset Sp(32, \bR)\subset SL(32, \bR)\subset GL(32, \bR)~.
\label{subgr}
\ee
The product of two spinor representations can be decomposed
as
$$
\Delta\otimes \Delta=\sum_{n=0}^5 \Lambda^n (\bR^{10,1})~.
$$
This is equivalent to the fact that a bi-spinor $\bar\chi\otimes \psi$
can be written as
\bea
\bar \chi\otimes \psi={1\ov 32}
\{(\bar\chi\psi) 1_{32}+\G^A(\bar\chi\G_{A}\psi)
+\dots
+{1\over 5!}\G^{A_1\dots A_5}(\bar\chi\G_{A_1\dots A_5}\psi) \}.
\label{frieza}
\eea
where $\bar\chi=\chi^tC^{-1}$ and $\{\Gamma^A; A=0, \dots 10\}$
is a basis of gamma matrices. This basis can be chosen
such that $\Gamma^0$ is anti-hermitian and the rest are hermitian.
Since the representation is real, all the gamma matrices
can be taken to be real
and so $\Gamma^0$ is skew-symmetric while the rest are symmetric
matrices. All skew products of gamma matrices are traceless.
So taking the trace in (\ref{frieza}), we get an identity. Since
a bi-spinor can be viewed as an element in $M_{32}(\bR)$,
the right-hand-side of (\ref{frieza}) takes values in $M_{32}(\bR)$.

To relate the Killing spinor equations to the
parallel transport equations of forms constructed from bi-spinors,
we need a relation between spinors and forms. For this, we choose
 a spinor $\epsilon\not=0$. Then setting $\chi=\epsilon$ in
(\ref{frieza}), we can define a map $i_\epsilon$
from $\Delta$ into $\Lambda^*(\bR^{10,1})$ by setting
\be
i_\epsilon(\psi)=
\bar\epsilon\otimes \psi~,
\label{incl}
\ee
 i.e. $(i_\epsilon(\psi))^\a{}_\b =(\epsilon^t C^{-1})^\a \psi_\b$.
We now prove that this map is $1-1$. Since the
map $i_\epsilon$ is linear, we have to show that its kernel vanishes.
Indeed, suppose that there is a $\psi\not=0$ in the kernel of the linear map.
This implies  that
$\chi_\a \epsilon_\b=0$ for all $\a, \b$. However
since neither $\psi$ nor $\epsilon$  vanish, they
must have at least one non-vanishing component each.
Thus there is at least a pair
$(\a,\b)$ such that $\chi_\a \epsilon_\b\not=0$.
Therefore the Kernel contains only the
zero element and the map is $1-1$.
This result implies that if there is a distinguished non-vanishing spinor,
we can  describe any other spinor in terms of forms which in the case
of eleven-dimensional supergravity have rank from zero to five. Therefore given
a non-vanishing spinor $\epsilon$,
$\Delta$  can be viewed as a subspace of $\sum_{n=0}^5\oplus \Lambda^n(\bR^{10,1})$.

\subsection{The holonomy of the supercovariant connection and $N$ Killing spinors}

The spinor bundle $S$ can be viewed as an associated vector bundle
of principal bundles for any of the groups in (\ref{subgr}), i.e.
$S=P(G)\times_\rho \Delta$
for $G=Spin(10,1), Sp(32, \bR), SL(32, \bR)$ or $GL(32, \bR)$. The representation $\rho$
 is the standard representation of
$GL(32, \bR)$ on $\Delta=\bR^{32}$ restricted on the subgroups $G$.

It turns out that for the investigation of Killing spinors in the context
of eleven-dimensional supergravity, it is most convenient to view $S$ as
an associated bundle of $SL(32, \bR)$. This is because the (reduced) holonomy
of the supercovariant derivative ${\cal D}$
 is contained in $SL(32, \bR)$, ${\rm hol}({\cal D})
\subseteq SL(32, \bR)$.
The supercovariant derivative of eleven-dimensional supergravity \cite{julia}
is
\be
{\cal D}_M\epsilon= \nabla_M \epsilon+ \Omega_M \epsilon
\label{supcon}
\ee
where
\be
\Omega_M=-{1\over 288} (\Gamma_M{}^{PQRS}F_{PQRS}-8 F_{MPQR} \Gamma^{PQR})~,
\label{supconb}
\ee
$\nabla_M$ is the spin covariant derivative induced from the Levi-Civita connection
and $M,N$,
$P,Q$,
$R$,
$S=0, \dots, 10$ are spacetime indices.
The curvature of the supercovariant derivative, ${\cal R}$,
is defined as,
\be
{\cal R}_{MN}=[{\cal D}_M, {\cal D}_N]~
\ee
and it can be expanded in a basis
of gamma matrices as
\be
{\cal R}_{MN}=\sum_{n=1}^5\phi_{MN}{}^{A_1\dots A_n} \Gamma_{A_1\dots A_n}~,
\ee
where $A_1, A_2, \dots, A_n=0,1,2, \dots, 10$ are frame indices
of eleven-dimensional spacetime.
The coefficients $\phi$ are functions of the bosonic fields of
eleven-dimensional supergravity $(g, F)$. An explicit expression
for these coefficients can be found in \cite{nicolai, gpjfof}.
The Lie algebra of the  holonomy group of a connection $D$ is determined by the
span of the values of the curvature $R_D(X,Y)$ evaluated on
any two vector fields $X,Y$.
The supercovariant curvature takes values in $\Delta\otimes \Delta$ but
it {\sl does not} contain any term which is {\sl zeroth order}
in gamma matrices, i.e. proportional to $1_{32}$.
Thus the trace
$$
{\rm tr}\left({\cal R}(X,Y)\right)=0
$$
on the spinor indices
vanishes. Therefore  ${\cal R}(X,Y)$  takes
values in the subset $M^0_{32}(\bR)\subset M_{32}(\bR)$ of
$32\times 32$ real matrices  which have vanishing
trace. The latter can be identified with the Lie algebra
 $sl(32, \bR)=M^0_{32}(\bR)$ of $SL(32,\bR)$.
Thus the {\sl reduced holonomy group} of the supercovariant derivative
is contained in $SL(32, \bR)$, ${\rm hol}({\cal D})\subseteq SL(32, \bR)$.
For a different proof of this see \cite{hull}.
In what follows, we shall consider $S=P(SL(32, \bR))\times_\rho \Delta$.
The supercovariant connection and supercovariant curvature are not
hermitian elements of the Clifford algebra.

The existence of parallel (Killing) spinors with respect to the
supercovariant derivative ${\cal D}$, i.e. spinors $\epsilon$ such that
$$
{\cal D}_M\epsilon=0~,
$$
implies
$$
{\cal R}_{MN}\epsilon=0
$$
and the holonomy group reduces  to a subgroup of $SL(32, \bR)$.
We shall identify
the subgroup of $SL(32, \bR)$ which allows the presence of $N$ Killing spinors.
This is equivalent to finding the subgroup of $SL(32, \bR)$ which is the
stability subgroup of $N$ spinors in $\Delta=\bR^{32}$. The reduction
of the holonomy group of the Levi-Civita connection
due to the existence of one Killing spinor  has been investigated in \cite{bryant, jose}.
The group $SL(32, \bR)$ acting with the standard representation
on $\Delta=\bR^{32}$ has two orbits. One is the origin $\{0\}$  of $\bR^{32}$
and the other is $\bR^{32}-\{0\}$. The stability subgroup of a non-vanishing
spinor is
$$
SL(31, \bR)\st \bR^{31}~.
$$
In matrix notation, any
$A\in SL(31, \bR)\st \bR^{31}$ can be written as
\be
A=\pmatrix{1& u^t\cr 0& B}
\ee
where $u\in \bR^{31}$ and $B\in SL(30, \bR)$.
Using this, it is easy to show that
 $SL(32, \bR)/SL(31, \bR)\st \bR^{31}=\bR^{32}-\{0\}$.
Next suppose that there are two linearly independent  Killing spinors $\epsilon_1$
and $\epsilon_2$.
Without loss of generality, we can choose the two spinors to
be
\be
\epsilon_1=e_1~~~~~~~~
\epsilon_2=\sum_{i=1}^{32} v^i e_i~,
\ee
where $\{e_i; i=1,2,\dots, 32\}$ is the standard basis in $\bR^{32}$.
Since $\epsilon_1$ and $\epsilon_2$ are linearly independent,
one of the components $v_2, \dots, v_{32}$ of $\epsilon_2$ must
be non-vanishing. Since  $SL(31, \bR)$ acts transitively on $\bR^{31}-\{0\}$,
we can use the stability subgroup of $\epsilon_1$ to set
$v_3=v_4=\dots =v_{32}=0$. It is then straightforward
to see that the stability subgroup of both $\epsilon_1$ and $\epsilon_2$
is $SL(30, \bR)\st (\bR^{30}\oplus \bR^{30})$. The group
$SL(30, \bR)$ acts with the fundamental representation on
both vector spaces in the direct sum. In matrix representation, any
$A\in SL(30, \bR)\st (\bR^{30}\oplus \bR^{30})$ can be written as
\be
A=\pmatrix{1& 0 &u_1^t\cr 0&1&u_2^t\cr 0&0&B}
\ee
where $u_1, u_2\in \bR^{30}$ and $B\in SL(30, \bR)$.
The orbits of $SL(31, \bR)\st \bR^{31}$ in $\bR^{32}$ are either the points
of the line $\bR(e_1)$ along the direction of the $e_1$ axis, or $\bR^{32}$
with the line along $e_1$ removed. I.e.
\bea
{\cal O}_x&=&\{x \in \bR(e_1)\}
\cr
{\cal O}'&=&\bR^{32}-\bR (e_1)~.
\eea
The stability subgroup  of a spinor in $\bR^{32}-\bR (e_1)$ is
$SL(30, \bR)\st (\bR^{30}\oplus \bR^{30})$ and so
\be
SL(31, \bR)\st \bR^{31}/SL(30, \bR)\st (\bR^{30}\oplus \bR^{30})=\bR^{32}-\bR (e_1)~.
\ee
Continuing in the same fashion,
the stability subgroup of $N$ linearly independent
Killing spinors is
\be
SL(32-N, \bR)\st (\oplus^N \bR^{32-N})~.
\ee
In matrix representation, an element $A\in SL(32-N, \bR)\st (\oplus^N \bR^{32-N})$
can be written as
\be
A=\pmatrix{1&0&0&\dots&0&0&u_1^t\cr  0&1&0&\dots&0&0&u_2^t\cr &&\vdots&&\vdots\cr
           0&0&0&\dots&0&1&u_k^t\cr 0&0&0&\dots&0&0&B}~,
\ee
where $u_1, \dots, u_N\in \bR^{32-N}$ and $B\in SL(32-N,\bR)$.
The orbits of $SL(33-N, \bR)\st (\oplus^{N-1} \bR^{33-N})$ in $\bR^{32}$ are as follows:
\bea
{\cal O}_x&=&\{x \in \bR(e_1, \dots, e_{k-1})\}
\cr
{\cal O}'&=&\bR^{32}-\bR(e_1, \dots, e_{k-1})~.
\eea
The stability subgroup of an element in $\bR^{32}-\bR(e_1, \dots, e_{N-1})$ is
$SL(32-N, \bR)\st (\oplus^{N} \bR^{32-N})$. Thus we have
\be
SL(33-N, \bR)\st (\oplus^{N-1} \bR^{33-N})/ SL(32-N, \bR)\st (\oplus^{N} \bR^{32-N})=
\bR^{32}-\bR(e_1,\dots, e_{N-1})~.
\label{ccc}
\ee

To  summarize, we have shown that the holonomy of the supercovariant connection
associated with a spacetime which admits
 $N$ Killing spinors is a subgroup of $SL(32-N, \bR)\st (\oplus^N \bR^{32-N})$.
 This result has also been stated in \cite{hull}, here we have given the proof
 in detail.
 For the spacetime to admit precisely $N$ Killing spinors the holonomy
 of the supercovariant connection must satisfy
 $$
SL(31-N, \bR)\st (\oplus^{N+1} \bR^{31-N})
\nsupseteq {\rm hol}({\cal D})\subseteq SL(32-N, \bR)\st (\oplus^N \bR^{32-N})
~.
$$

It is clear from the discussion in this section that at every point of
spacetime, $SL(32, \bR)$ acts transitively
on the coset $SL(32, \bR)/SL(32-N, \bR)\st (\oplus^N \bR^{32-N})$.
The latter can be thought of as the space of inequivalent configurations
with N Killing spinors\footnote{This is similar to the statement that
$GL(2n, \bR)/GL(n, \bC)$ parameterizes the space of complex structures
at a point.}. However as we shall see $SL(32, \bR)$ does not preserve
the space of supersymmetric planar probe M-brane configurations.
It would be of interest to see whether this difference of behaviour
between the $SL(32, \bR)$ and $SO(32, \bR)$ groups
has applications in the context of symmetries of M-theory.
It is the latter group that acts on the M-brane charges.

\subsection{Supercovariant curvature and $N$ Killing spinors}

As we have seen the necessary and sufficient condition for
the existence of $N$ Killing spinors is the reduction of the
holonomy of the supercovariant derivative ${\cal D}$ from
$SL(32, \bR)$ to the $SL(32-N, \bR)\st (\oplus^N \bR^{32-N})$ subgroup.
However, this may be seen as rather implicit so we shall 
express this condition in terms of
relations for the metric $g$ and the four-form field
strength $F$ of eleven-dimensional supergravity.
Let $\epsilon$
be a Killing spinor, then we have seen that
$\epsilon$ is a eigenvector of ${\cal R}_{MN}$ with zero eigenvalue
\be
{\cal R}_{MN}\epsilon=0~.
\label{curvan}
\ee
For simply connected spacetimes (\ref{curvan}) is also
a sufficient condition\footnote{
We assume that the spacetime is simply connected. If it is not,
we take the universal cover.}.
Since $M_{32}^0=sl(32, \bR)$, we can express the supercovariant
curvature in
terms of a basis
$\{{\rm m}_{ab}; a,b=1, \dots, 32\}$ basis of $gl(32, \bR)$. I.e.
\be
{\cal R}_{MN}=\sum^5_{n=1} \phi_{MN}{}^{A_1A_2\dots A_n} \Gamma_{A_1A_2\dots A_n}=\sum_{n=1}^5
\phi_{MN}^{A_1A_2\dots A_n} X_{A_1\dots A_n}^{ab} {\rm m}_{ab}={\cal R}^{ab}_{MN} {\rm m}_{ab} ~,
\ee
where $X$ is the transformation from the basis of
gamma matrices basis to the basis
of $gl(32,\bR)$, see appendix A.
We can adapt an $SL(32, \bR)$ moving
frame in spacetime, such that  $\epsilon$ is along the
$e_1$ axis. Using the basis $m_{ab}$ defined in
appendix A,
the conditions on the supercovariant curvature for a spacetime
to admit one Killing spinor are
\be
{\cal R}^{a1}_{MN}=0~, ~~~a=1, \dots, 32.
\ee
Similarly, the conditions for  spacetime to admit $N$ Killing spinors
are
\be
{\cal R}^{ab}_{MN}=0~,~~~a=1, \dots, 32~, ~~ b=1, \dots, N~.
\ee
These  together with the field equations of eleven-dimensional
supergravity is the {\sl full set} of conditions that  a spacetime should
satisfy
in order for it to admit
$N$ Killing spinors. This extends the first part of the programme for the
classification of maximal supersymmetric solutions
\cite{gpjfof} to backgrounds that preserve less than maximal supersymmetry.
The conditions we have derived, although explicit, are rather involved and it remains to be
seen whether the second part of the programme of \cite{gpjfof}
can be carried over to the $N<32$ case as well.

\newsection{Topological Aspects of Spacetimes with $N$ Killing spinors}

There may be topological conditions for the existence of $N$ parallel
spinors in spacetime. Since the holonomy of the supercovariant
connection reduces, the structure group of the spinor bundle
reduces as well to a subgroup of the holonomy group (see e.g.~ \cite{kn}).
In particular, the structure group of $S$ reduces from $SL(32, \bR)$
to the subgroup $SL(32-N, \bR)\st (\oplus^N \bR^{32-N})$. The topological
conditions for the existence of
$N$ Killing spinors are identified with the obstructions to the above reduction
of the structure group.
It is known that there
are obstructions in the reduction of the structure group
of the spin bundle due to the existence of a no-where vanishing spinor
on two-, four- and eight-dimensional manifolds, see  \cite{isham, ishamb}
and references therein.
For manifolds with dimension
more than eight there is no topological obstruction because the
rank of the spin bundle is much greater than the dimension of the manifold.
   In particular
in eleven-dimensions there is no obstruction for the
existence of a no-where vanishing spinor.
However, the reduction
of the structure group from $SL(32, \bR)$ to $SL(32-N, \bR)\st (\oplus^N \bR^{32-N})$
imposes
some topological conditions on
spacetime for large enough $N$, as we now show.

It is well-known that the structure group of $S$  reduces
from $SL(32, \bR)$ to $SL(32-N, \bR)\st (\oplus^N \bR^{32-N})$ iff the associated
bundle $P(SL(32, \bR))\times_{\bar \rho} Z$,
  with fibre $Z$ the coset space
\be
Z=SL(32, \bR)/SL(32-N, \bR)\st (\oplus^N \bR^{32-N})~,
\label{klip}
\ee
admits a global section.
In (\ref{klip}), $\bar \rho$ is the left action of $SL(32, \bR)$ on $Z$.
There may be  obstructions to the existence
of such a global section. These obstructions are elements in the
cohomology groups of spacetime\footnote{We assume that the spacetime is homotopic
to a CW complex.}. The first obstruction class, called primary, lies in the
first non-vanishing cohomology group $H^n(M, \pi_{n-1}(Z))$. If the primary
obstruction vanishes, there may be secondary obstructions which lie
in higher cohomology groups (see e.g. \cite{steenrod}).
In our case, in order to
find the cohomology group of spacetime in which
the primary  obstruction
lies, observe that the  coset space $Z$ is homotopic
to the Stiefel manifold $V_{32, N}=SO(32,\bR)/SO(32-N,\bR)$. This can be seen
by observing that $SL(K, \bR)$ contracts to its  maximal compact subgroup
 $SO(K)$.
The Stiefel manifold $V(K,N)=SO(K)/SO(K-N)$ is $K-N-1$ connected. This means
that
$$
\pi_r(V(K,N))=0~~~~~~~r=0, \dots, K-N-1~.
$$
In addition,
\bea
\pi_r (V(r+N, N))&=&\bZ~~~~~r\in 2\bZ
\cr
\pi_r (V(r+N, N))&=&\bZ_2~~~~~ r\in 2\bZ+1~, ~~N\not=1
\cr
\pi_r(V(r+1,1)&=&\bZ~~~~~r\in \bZ
\eea
Because $V_{32,N}$ is $31-N$ connected,  the primary obstruction
lies in the  cohomology group
\be
H^{33-N}(M, \pi_{32-N}(V_{32,N}))~.
\label{prim}
\ee
If spacetime is homotopic to a point, the primary   and all the
 secondary obstructions  vanish and so there are no topological
conditions for the existence of any number of Killing spinors.
Obstructions may arise when the spacetime is topologically non-trivial.

{}For  one Killing spinor, $N=1$, in eleven-dimensional spacetime,
(\ref{prim}) always vanishes. So there is no primary obstruction. There are no
secondary obstructions either because they lie in cohomology groups of higher degree.
So there is no topological restriction on spacetime for a single Killing
spinor to exist.
 The same is true for $N<22$. For example there is no topological obstruction
for the existence of $N=16$ Killing spinors in eleven-dimensional spacetime.

If $N=22$ and $H^{11}(M)\not=0$,
there may be a non-vanishing primary obstruction which lies in
$H^{11}(M, \pi_{10} (V_{32, 22}))=H^{11}(M, \bZ)$.
To identify this obstruction, let us assume that we have
 $N$ Killing spinors, and  let $\Delta_{N}$ be the
$SL(32-N, \bR)\st (\oplus^{N} \bR^{32-N})$-invariant subspace in $\Delta$.
We have that
\be
0\rightarrow \Delta_{N}\rightarrow \Delta\rightarrow \Delta/\Delta_{N}\rightarrow 0~.
\ee
Observe that this short exact sequence  does not split, i.e.
$\Delta$ cannot
be written as a direct sum of $\Delta_N$ and another space in such way that
the sum is preserved by the $SL(32-N, \bR)\st (\oplus^N \bR^{32-N})$ group action.
Using this exact sequence of vector spaces, we can define the bundles
$S_N=M\times \Delta_{N}$,
$S$ and
$S_{\Delta/\Delta_N}=P(SL(32-N, \bR)\st
(\oplus^{N} \bR^{32-N}))\times_\rho \Delta/\Delta_{N}$,
where $S$ is the spin bundle, $\rho$ is the factor representation of
$SL(32-N, \bR)\st (\oplus^{N} \bR^{32-N})$ on $\Delta/\Delta_{N}$ and $S_N$
is the bundle of Killing spinors.
Since there is no topological obstruction for the existence of $N=21$ Killing spinors,
we can assume that the spacetime admits $N=21$ Killing spinors and so we can define
the bundle $S_{\Delta/\Delta_{21}}$. The obstruction for the existence of
$N=22$ Killing spinors is the Euler class of the bundle $S_{\Delta/\Delta_{21}}$.
(Observe this bundle has the same rank as the dimension of spacetime.)
If this Euler class vanishes, there are no further
secondary obstructions.

\newsection{Killing spinors and forms}

The supercovariant derivative ${\cal D}$, as any other covariant derivative
of the spin bundle $S$, can be extended to a covariant derivative
of $S\otimes S$ as ${\cal D}\otimes 1+1\otimes {\cal D}$.
But $S\otimes S$ can be thought of as a sub-bundle of $\Lambda^*(M)$. So
${\cal D}$ can be extended to a covariant derivative on the space of forms.
In particular, we have
\bea
{\cal D}_M (\chi_\a \psi_\b)&=&{\cal D}_M \chi_\a \psi_\b+ \chi_\a {\cal D}_M\psi_\b
\cr
&=&
{1\ov 32}[
-C_{\a\b} {\cal D}_M\bar\chi\psi+\G^N_{\a\b} {\cal D}_M\bar\chi\G_{N}\psi
+\dots
+{1\over 5!}\G^{N_1\dots N_5}_{\a\b} {\cal D}_M\bar\chi\G_{N_1\dots N_5}\psi
\cr
&-&
C_{\a\b}\bar\chi {\cal D}_M\psi+\G^N_{\a\b}\bar\chi\G_{N}{\cal D}_M\psi
+\dots
+{1\over 5!}\G^{N_1\dots N_5}_{\a\b}\bar\chi\G_{N_1\dots N_5} {\cal D}_M\psi ]
\eea
Using (\ref{supcon}), we can rewrite the above
expression as
\bea
{\cal D}_M (\chi_\a \psi_\b)&=&{1\over32} [-C_{\a\b}{\nabla}_M(\bar\chi\psi)
 +\G^N_{\a\b}\nabla(\bar\chi\G_{N}\psi)
+\dots
+{1\over 5!}\G^{N_1\dots N_5}_{\a\b}{\nabla}_M(\chi\G_{N_1\dots N_5}\psi)
\cr
&&
-C_{\a\b}\bar\chi (\bar \Omega_M +\Omega_M)\psi+\G^N_{\a\b}\bar\chi
(\bar \Omega_M\G_{N}+\G_N \Omega_M)\psi
+\dots
\cr
&&
+{1\over 5!}
\G^{N_1\dots N_5}_{\a\b}\bar\chi(\bar \Omega_M\G_{N_1\dots N_5}+\G_{N_1\dots N_5}\Omega_M)\psi ]~.
\label{cona}
\eea
The supercovariant derivative of the k-form
$\bar\chi \Gamma_{N_1\dots N_k}\psi$
associated with the bi-spinor
$\chi\otimes \psi$ is
\be
{\cal D}_M (\bar\chi \Gamma_{N_1\dots N_k}\psi)= \nabla_M (\bar\chi \Gamma_{N_1\dots N_k}\psi)
+\bar\chi(\bar \Omega_M\G_{N_1\dots N_k}+\G_{N_1\dots N_k}\Omega_M)\psi~.
\label{covb}
\ee
The second part of the right-hand-side of (\ref{covb})
can be rewritten in terms of forms  associated with the bi-spinor
$\chi\otimes \psi$ contracted with $F$. For a generic four-form field strength $F$,
${\cal D}$ does not preserve the degree of the form that
it acts on. In other words, when
the connection is evaluated on a one-form, the term involving $\Omega_M$ mixes
the forms of various degrees.

Suppose now that $\chi=\epsilon_i$ and $\psi=\epsilon_j$
are Killing spinors, so that
${\cal D}_M (\epsilon_i\otimes \epsilon_j)=0$. This
implies that the forms associated with the bi-spinors $\epsilon_i\otimes \epsilon_j$
are parallel with respect to the superconnection ${\cal D}$.
 Let us denote the
forms associated with the bi-spinor $\epsilon_i\otimes \epsilon_j$
collectively by $\tau_{ij}$. Then we have that
\be
{\cal D}_M\tau_{ij}=0~, ~~~~~~ i,j=1,\dots, N~.
\label{partau}
\ee
 These conditions are explicitly given  in
\cite{pakis} after adjusting for conventions ($F\rightarrow -F$).

Here, we shall show that  only $N$
of the conditions
(\ref{partau}) are independent, as many
as the number of Killing spinors. In addition, we shall show that these
$N$ independent conditions are equivalent to the Killing spinor equations.
To show this recall that if there is a non-vanishing spinor $\epsilon$, one can construct
an inclusion of $\Delta$ in $\Lambda^*(\bR^{10,1})$ as in (\ref{incl}).
Setting $\epsilon=\epsilon_1$, the independent conditions are equivalent
to $\epsilon_1\otimes \epsilon_i$ for $i=1, \dots, N$ being parallel with
respect to ${\cal D}$. Therefore the independent conditions are
\be
{\cal D}_M \tau_{1i}=0~, ~~~~~~i=1,\dots,N~.
\label{kfcon}
\ee
Conversely, if (\ref{kfcon}) is satisfied, then ${\cal D}_M\epsilon_i=0$.
To show this  consider the spinor $\epsilon=\epsilon_1\not=0$ and  suppose that
${\cal D}\epsilon\not=0$ but ${\cal D} \tau_{11}=0$. In this case we have
\be
{\cal D}\tau_{11}=0={\cal D}\epsilon\otimes \epsilon+ \epsilon\otimes {\cal D}\epsilon~.
\ee
It is easy to see that if there is a component of ${\cal D}\epsilon$ which does
vanish, then $\epsilon=0$ which is a contradiction.
Next we take
\be
0={\cal D}\tau_{1i}={\cal D}(\epsilon\otimes \epsilon_i)=\epsilon\otimes {\cal D}\epsilon_i~,
\ee
which implies that ${\cal D}\epsilon_i=0$. This
completes the proof that the Killing spinor
equations are equivalent to (\ref{kfcon}).

\section{Symmetry Superalgebra}

The bosonic symmetries of a supersymmetric  background of a supergravity theory
 are generated by Killing vectors
 which in addition leave the remaining form field-strengths invariant.
 There are two types of such Killing vectors  depending
on whether or not they can be constructed from Killing spinors.
The Killing vectors
\be
K_{ij}=K_{ji}=\bar\epsilon_i\Gamma^M\epsilon_j \partial_M~, ~~~~i,j=1, \dots, N
\label{kvs}
\ee
associated with the Killing spinors $\epsilon_i, \epsilon_j$
are not all independent.
 The maximal number of linearly independent Killing vectors $K_{ij}$ is
  ${1\over2} N(N+1)$.
For $N>11$, this number is larger than the maximal number
of linearly independent Killing vectors allowed on
 eleven-dimensional spacetime.
In most known supersymmetric M-theory backgrounds, the number
of $K_{ij}$  Killing vectors  is further restricted. For example
pp-wave backgrounds associated with  multi-centre harmonic functions
 admit a single null Killing vector despite
the fact that they preserve sixteen supersymmetries, $N=16$.

Let ${\cal G}={\cal G}_0\oplus {\cal G}_1$ be the symmetry
superalgebra of an M-theory background
--${\cal G}_0$ and ${\cal G}_1$ are the even and odd parts, respectively.
In addition, we introduce a basis $\{Q_i: i=1,\dots,N\}$ of  ${\cal G}_1$.
The even part ${\cal G}_0$ is spanned
by  $\{T_{ij}: i,j=1,\dots, N\}$
 associated with Killing vectors (\ref{kvs})
 and by $\{T_a: a=1, \dots, I\}$ which are the
even generators associated with the rest of the Killing vectors of the
  M-theory background.
The main aim of this section is to show that the
symmetry superalgebra of a supersymmetric M-theory  background can be written as
\bea
[T_{ij}, Q_k]&=& f_{ij,k}{}^l Q_l
\cr
[T_a, Q_i]&=& f_{a i}{}^l Q_l
\cr
[T_{ij}, T_{mn}]&=& f_{ij,mn}{}^{kl} T_{kl}
\cr
[T_a, T_{ij}]&=& f_{a i}{}^k T_{kj}+ f_{a j}{}^k T_{ik}
\cr
[T_a, T_b]&= &f_{ab}{}^c T_c+f_{ab}{}^{ij} T_{ij}
\cr
\{Q_i, Q_j\}&=&T_{ij}~.
\label{symalg}
\eea
The $f$'s above are structure constants obeying
\be
f_{ij,mn}{}^{kl}= f_{ij,m}{}^{(k}\delta_n^{l)}
+ f_{ij,n}{}^{(k}\delta_m^{l)}
\label{russell}
\ee
and
\be
f_{ij,k}{}^p=h_{ki}\delta_j^p+h_{kj}\delta_i^p~,
\label{wittgenstein}
\ee
where $h_{ij}$ is a constant antisymmetric second-rank tensor, $h_{ij}=-h_{ji}$. Note that
(\ref{russell}),(\ref{wittgenstein}) imply that the structures
constants $f_{ij,mn}{}^{kl}$ are antisymmetric under
$(m,n)\leftrightarrow (i,j)$.
In (\ref{symalg}) we have treated all the $T_{ij}$ generators as
independent. However as we have mentioned this is not always the case
because the associated Killing vectors $K_{ij}$ may be
linearly dependent. If the $T_{ij}$s are linearly dependent, then the constant
anti-symmetric matrix $h$ should be appropriately restricted. The generators
$T_{ij}$ span a subalgebra of ${\cal G}_0$ which we denote with ${\cal G}^s_0$

We shall first show closure. It is well known that the algebra
of Killing vectors on a manifold closes. This implies the closure
for all the even generators. The commutator of even with
odd generators can be computed from the spinorial Lie
derivative of the Killing spinors along the direction of the
Killing vector fields \cite{pktc, josev}. To show the closure of the commutator
of  even with odd generators, we have to show that the spinorial
Lie derivative of a Killing spinor with respect to any Killing vector
which is a symmetry of the background is again a Killing spinor.
 Indeed, let $K$ be any vector which satisfies
\be
{\cal L}_K g=0~~~~~~~{\cal L}_KF=0
\label{kmf}
\ee
and $\epsilon$ be a Killing spinor. The spinorial Lie
derivative of $\epsilon$ along $K$ is
\be
{\cal L}_K\epsilon= K^M \nabla_M\epsilon
+{1\over4} \nabla_MK_N \Gamma^{MN}\epsilon~.
\label{slie}
\ee
 Using (\ref{kmf}) and the
properties of the spinorial Lie derivative, we find that
\be
[{\cal L}_K, {\cal D}]=0~.
\ee
So if ${\cal D}\epsilon=0$, then ${\cal D}{\cal L}_K\epsilon=0$
and the statement is shown.

Closure of the commutator of even with odd generators
  implies that there are (structure) constants
$f$ such that
\be
{\cal L}_{K_{ij}} \epsilon_k=  f_{ij,k}{}^{l} \epsilon_l~.
\label{lks}
\ee
To show that $f$ is constant, it suffices to take the supercovariant
derivative on both sides of (\ref{lks}). Using (\ref{lks}),
we can determine the structure constants associated
with the Lie bracket of the Killing vectors,
\bea
[K_{ij}, K_{mn}]&=&{\cal L}_{K_{ij}}(\bar{\epsilon}_m\Gamma \epsilon_n )
\nn\\
&=&f_{ij,m}{}^\ell K_{\ell n}
+ f^{ij,n}{}^\ell K_{\ell m} ~.
\label{lkk}
\eea
This establishes (\ref{russell}) and
the commutator of the even generators associated
with Killing vector fields constructed from Killing spinors.

Closure also implies that there are constants such that
\be
{\cal L}_{K_a} \epsilon_i= f_{ai}{}^j \epsilon_j~.
\ee
One can then easily determine the structure constants of
 the commutator $[T_a, T_{ij}]$ from those above.
We have thus
shown that most of the structure constants
are determined from the brackets
of the even generators with the odd generators.

It may seem natural to take $f_{ab}{}^{ij}=0$ in (\ref{symalg})so that
the
even $\{T_a\}$ generators
act on the remaining algebra as external automorphisms. In many known
examples, like the maximal supersymmetric plane-waves
(\cite{kg, gpja, gpjb}), it holds that $f_{ab}{}^{ij}=0$.
However, it is not apparent
that these structure constants will vanish in general.

The commutator of the odd with odd generators can be derived from the
squaring of two Killing spinors. Therefore we
have
\be
\{ Q_i, Q_j\}=T_{ij}~.
\ee
(See appendix B for a further discussion.)
In order to show (\ref{wittgenstein}), we first note that
\be
f_{(ij,k)}{}^{l}=0~.
\label{alle}
\ee
Indeed, let us examine (\ref{lks}) more closely. The Lie derivative of
a spinor transforms  as the symmetrized tensor
product of two spinors tensored with a four-form and a spinor,
\be
{\cal L}_{K_{ij}} \epsilon_k \sim
(00001)^{2\otimes_s}\otimes(00010)\otimes(00001)
\sim 10(00001)\oplus \dots ~,
\ee
where we have noted that there are ten spinors in the decomposition.
These can be given  explicitly,
\bea
S_{ij,k}^{(1)}&:=&
{1\over 3!}(\Gamma_{N_1N_2N_3}\epsilon_k)(K_{ij})_{N_4}F^{N_1\dots N_4}\nn\\
S_{ij,k}^{(2)}&:=&
{1\over 4!}(\Gamma_{N_1\dots N_5}\epsilon_k)(K_{ij})^{N_1}F^{N_2\dots N_5}\nn\\
S_{ij,k}^{(3)}&:=&
{1\over 4}(\Gamma_{N_1N_2}\epsilon_k)
(\Omega_{ij})_{N_3N_4}F^{N_1\dots N_4}\nn\\
S_{ij,k}^{(4)}&:=&
{1\over 4!}(\Gamma_{N_1\dots N_4}
\epsilon_k)(\Omega_{ij})_{N_5}{}^{N_1}F^{N_2\dots N_5}\nn\\
S_{ij,k}^{(5)}&:=&
{1\over 2. 4!}(\Gamma_{N_1\dots N_6}
\epsilon_k)(\Omega_{ij})^{N_1N_2}F^{N_3\dots N_6}\nn\\
S_{ij,k}^{(6)}&:=&
{1\over 4!}(\Gamma_{N_1}
\epsilon_k)(\Sigma_{ij})_{N_2\dots N_5}{}^{N_1}F^{N_2\dots N_5}\nn\\
S_{ij,k}^{(7)}&:=&
{1\over 2. 3!}(\Gamma_{N_1N_2N_3}
\epsilon_k)(\Sigma_{ij})_{N_4N_5N_6}{}^{N_1N_2}F^{N_3\dots N_6}\nn\\
S_{ij,k}^{(8)}&:=&
{1\over 4. 3!}(\Gamma_{N_1\dots N_5}
\epsilon_k)(\Sigma_{ij})_{N_6N_7}{}^{N_1N_2N_3}F^{N_4\dots N_7}\nn\\
S_{ij,k}^{(9)}&:=&
{1\over 4! 3!}(\Gamma_{N_1\dots N_7}
\epsilon_k)(\Sigma_{ij})_{N_8}{}^{N_1\dots N_4}F^{N_5\dots N_8}\nn\\
S_{ij,k}^{(10)}&:=&
{1\over 5! 4!}(\Gamma_{N_1\dots N_9}
\epsilon_k)(\Sigma_{ij})^{N_1\dots N_5}F^{N_6\dots N_9} ~.
\label{strcts}
\eea
Since the structures above form a basis,
the Lie derivative can be expressed in terms of them. Indeed
from (\ref{slie}), (\ref{strcts}) and
using the Killing property of $\epsilon_k$, one finds,
\be
{\cal L}_{K_{ij}} \epsilon_k={1\over 6}S_{ij,k}^{(1)}
+{1\over 12}S_{ij,k}^{(2)}
-{1\over 3}S_{ij,k}^{(3)} +10S_{ij,k}^{(10)}~.
\label{blum}
\ee
By Fierz-rearranging (\ref{strcts}) we get the following relations
\bea
S_{k(i,j)}^{(1)}&=&-{1\over 16}S_{ij,k}^{(1)}-{1\over 8}S_{ij,k}^{(2)}
+{1\over 2}S_{ij,k}^{(4)}
-{1\over 4}S_{ij,k}^{(5)}+15S_{ij,k}^{(6)}\nn\\
&&-{15\over2}S_{ij,k}^{(7)}-
{15\over2}S_{ij,k}^{(9)}-15S_{ij,k}^{(10)}\nn\\
S_{k(i,j)}^{(2)}&=&-{7\over32}S_{ij,k}^{(1)}-{5\over32}S_{ij,k}^{(2)}
+{7\over16}S_{ij,k}^{(3)}+{5\over4}S_{ij,k}^{(4)}
-{3\over16}S_{ij,k}^{(5)}-{75\over4}S_{ij,k}^{(6)}\nn\\
&&+{45\over4}S_{ij,k}^{(7)}+{15\over4}S_{ij,k}^{(8)}
+{15\over4}S_{ij,k}^{(9)}+{45\over4}S_{ij,k}^{(10)}\nn\\
2S_{k(i,j)}^{(3)}&=&{3\over16}S_{ij,k}^{(2)}
-{1\over8}S_{ij,k}^{(3)}
-{3\over8}S_{ij,k}^{(5)}+{45\over2}S_{ij,k}^{(6)}
+{15\over2}S_{ij,k}^{(8)}
+{45\over2}S_{ij,k}^{(10)}\nn\\
2S_{k(i,j)}^{(4)}&=&{7\over64}S_{ij,k}^{(1)}+{5\over32}S_{ij,k}^{(2)}
+{5\over8}S_{ij,k}^{(4)}
-{3\over16}S_{ij,k}^{(5)}-{75\over4}S_{ij,k}^{(6)}
+{45\over8}S_{ij,k}^{(7)}\nn\\
&&-{15\over8}S_{ij,k}^{(9)}
-{45\over4}S_{ij,k}^{(10)}\nn\\
2S_{k(i,j)}^{(5)}&=&-{21\over32}S_{ij,k}^{(1)}-{9\over32}S_{ij,k}^{(2)}
-{21\over16}S_{ij,k}^{(3)}-{9\over4}S_{ij,k}^{(4)}
+{1\over16}S_{ij,k}^{(5)}-{135\over4}S_{ij,k}^{(6)}\nn\\
&&+{15\over4}S_{ij,k}^{(7)}-{45\over4}S_{ij,k}^{(8)}
+{45\over4}S_{ij,k}^{(9)}-{15\over4}S_{ij,k}^{(10)}\nn\\
5!S_{k(i,j)}^{(6)}&=&{7\over32}S_{ij,k}^{(1)}-{5\over32}S_{ij,k}^{(2)}
+{7\over16}S_{ij,k}^{(3)}-{5\over4}S_{ij,k}^{(4)}
-{3\over16}S_{ij,k}^{(5)}-{75\over4}S_{ij,k}^{(6)}\nn\\
&&-{45\over4}S_{ij,k}^{(7)}+{15\over4}S_{ij,k}^{(8)}
-{15\over4}S_{ij,k}^{(9)}+{45\over4}S_{ij,k}^{(10)}\nn\\
5!S_{k(i,j)}^{(7)}&=&-{21\over16}S_{ij,k}^{(1)}+{9\over8}S_{ij,k}^{(2)}
+{9\over2}S_{ij,k}^{(4)}
+{1\over4}S_{ij,k}^{(5)}-135S_{ij,k}^{(6)}
-{15\over2}S_{ij,k}^{(7)}\nn\\
&&+{45\over2}S_{ij,k}^{(9)}+15S_{ij,k}^{(10)}\nn\\
5!S_{k(i,j)}^{(8)}&=&{15\over16}S_{ij,k}^{(2)}
+{35\over8}S_{ij,k}^{(3)}
-{15\over8}S_{ij,k}^{(5)}+{225\over2}S_{ij,k}^{(6)}
-{45\over2}S_{ij,k}^{(8)}
+{225\over2}S_{ij,k}^{(10)}\nn\\
5!S_{k(i,j)}^{(9)}&=&-{35\over16}S_{ij,k}^{(1)}+{5\over8}S_{ij,k}^{(2)}
-{5\over2}S_{ij,k}^{(4)}
+{5\over4}S_{ij,k}^{(5)}-75S_{ij,k}^{(6)}
+{75\over2}S_{ij,k}^{(7)}\nn\\
&&-{45\over2}S_{ij,k}^{(9)}+75S_{ij,k}^{(10)}\nn\\
5!S_{k(i,j)}^{(10)}&=&-{21\over32}S_{ij,k}^{(1)}+{9\over32}S_{ij,k}^{(2)}
+{21\over16}S_{ij,k}^{(3)}-{9\over4}S_{ij,k}^{(4)}
-{1\over16}S_{ij,k}^{(5)}+{135\over4}S_{ij,k}^{(6)}\nn\\
&&+{15\over4}S_{ij,k}^{(7)}+{45\over4}S_{ij,k}^{(8)}
+{45\over4}S_{ij,k}^{(9)}+{15\over4}S_{ij,k}^{(10)}~.
\label{plirps}
\eea
Symmetrizing all $i,j,k$ indices, (\ref{plirps})
reduces to a system of ten equations on ten unknowns
$S_{(ij,k)}^{(1)},\dots S_{(ij,k)}^{(10)}$. It turns out however
that only seven of the equations are linearly independent. Consequently,
all structures can be expressed in terms of three independent
ones which we may take to be $S_{(ij,k)}^{(1)},\dots S_{(ij,k)}^{(3)}$.
This is in accordance to the fact that there are exactly
three spinors in the decomposition of the tensor product of
a four-form and the symmetrized product of thee spinors,
\be
(00010)\otimes(00001)^{3\otimes_s}=3(00001)\oplus\dots~.
\ee
Explicitly we have,
\bea
S_{(ij,k)}^{(4)}&=&{1\over8}\left(3S_{(ij,k)}^{(1)}+4S_{(ij,k)}^{(2)}
-4S_{(ij,k)}^{(3)}\right)\nn\\
S_{(ij,k)}^{(5)}&=&{1\over2}\left(-2S_{(ij,k)}^{(1)}-S_{(ij,k)}^{(2)}
-2S_{(ij,k)}^{(3)}\right)\nn\\
S_{(ij,k)}^{(6)}&=&{1\over120}\left(-S_{(ij,k)}^{(2)}
 +2S_{(ij,k)}^{(3)}\right)\nn\\
S_{(ij,k)}^{(7)}&=&{1\over120}\left(-S_{(ij,k)}^{(1)}+4S_{(ij,k)}^{(2)}
-4S_{(ij,k)}^{(3)}\right)\nn\\
S_{(ij,k)}^{(8)}&=&{1\over12}S_{(ij,k)}^{(3)}\nn\\
S_{(ij,k)}^{(9)}&=&-{1\over24}S_{(ij,k)}^{(1)} \nn\\
S_{(ij,k)}^{(10)}&=&{1\over120}\left(-2S_{(ij,k)}^{(1)}-S_{(ij,k)}^{(2)}
+4S_{(ij,k)}^{(3)}\right) ~.
\label{above}
\eea
From (\ref{above}), (\ref{blum}) we conclude that
$$
{\cal L}_{K_{(ij}} \epsilon_{k)}=0,
$$
and therefore (\ref{alle}) is proven.

As a consequence, in the case where there is one Killing spinor
one cannot generate another Killing spinor by taking the Lie
derivative of the first one. In other words, one supersymmetry does
not necessarily imply a second one.

A direct calculation using
(\ref{slie}) and the Killing property of the spinors, reveals that
\bea
{\cal L}_{K_{ij}}(\bar{\epsilon}_m\Gamma_M \epsilon_n )&=&
{1\over 3} F_{MN_1N_2N_3} ~\bar \epsilon_m\Gamma^{N_1N_2}\epsilon_n~(K_{ij})^{N_3}\nn\\
&+&{1\over 6} ({}^*F)_{MN_1\dots N_6}~
\bar \epsilon_m\Gamma^{N_1\dots N_5}\epsilon_n (K_{ij})^{N_6}
-(m,n)\leftrightarrow (i,j)~,
\eea
which implies that the right-hand side of (\ref{lkk}) is
antisymmetric under $(m,n)\leftrightarrow (i,j)$ as of course it should.
Consequently, the structure constants obey
\be
f_{ij,m}{}^{(k}\delta_n^{l)}
+ f_{ij,n}{}^{(k}\delta_m^{l)}+(m,n)\leftrightarrow (i,j)=0~.
\label{chomsky}
\ee
Contracting (\ref{chomsky})
with $\delta_k^m\delta^n_l$ and using (\ref{alle}) we
obtain
\be
f_{ij,q}{}^q=0~.
\label{plato}
\ee
Finally, contracting (\ref{chomsky})
with $\delta^n_l$ and using (\ref{plato}),(\ref{alle}) we obtain
(\ref{wittgenstein}) with
\be
h_{ij}=-{1\over N+1}f_{iq,j}{}^q ~.
\ee
Note that $h_{ij}$ is antisymmetric by virtue of (\ref{alle}) and (\ref{plato}).

As a consequence of (\ref{wittgenstein}), the Jacobi identity for
the even generators $T_{ij}$, is automatically satisfied. Indeed,
(\ref{russell}),(\ref{wittgenstein}) imply that the commutators
of the subalgebra ${\cal G}_0^s$ are
\be
[T_{ij},T_{mn}]=4h_{(m|(i}T_{j)|n)}~,
\label{descartes}
\ee
from which it follows that
\be
[[T_{ij},T_{mn}],T_{kl}]+{\rm cyclic}=0~.
\ee

As an  example,
let us analyze
the algebra of isometries associated with Killing spinors,
in the case where there are
exactly two  \footnote{In the case
of a single Killing spinor there can be only one
associated Killing vector
and the algebra of isometries is isomorphic to $u(1)$.}
Killing spinors $\epsilon_1$,  $\epsilon_2$.

In view of (\ref{descartes}) we have the following algebra
\be
[K_{11},K_{22}]=-4\alpha K_{12}; ~~~~
{}[K_{11},K_{12}]=-2\alpha K_{11}; ~~~~
{}[K_{12},K_{22}]=-2\alpha K_{22}~,
\label{alae}
\ee
where we have set $\alpha:=h_{12}=-h_{21}$.

\noindent $\bullet$ If $\alpha=0$, the algebra
(\ref{alae}) is isomorphic
to $u(1), \oplus^2 u(1)$ or $\oplus^3 u(1)$, according
to the number of linearly independent Killing vectors.

\noindent $\bullet$ If $\alpha \neq 0$, we set
\be
X_1:=-{1\over 2\alpha}K_{12},
~~~~ X_2:={1\over 4\alpha}(K_{11}+K_{22}),
~~~~ X_3:={1\over 4\alpha}(-K_{11}+K_{22}),
\ee
so that
(\ref{alae}) takes the form
\be
[X_1,X_2]=X_3; ~~~~
{}[X_2,X_3]=X_1; ~~~~
{}[X_3,X_1]=-X_2~.
\label{euripides}
\ee
This algebra is isomorphic to $so(1,2;\bR)$.

More generally:

\noindent $\bullet$ In the case where the number of Killing spinors
is even ($N=2r$)
and $h_{ij}$
is non-degenerate,
the algebra ${\cal G}_0^s$ is
isomorphic to a real form  of $ sp(2r, \bC)$.
Perhaps the easiest way to see this is to note that
an explicit matrix realization of (\ref{descartes}) is given
by
\be
(T_{ij})^l{}_k=\delta^l_j h_{ki}+\delta^l_i h_{kj}~.
\ee
It is then straightforward to verify that
\be
(T_{ij}^{\rm t})_k{}^m h_{ml} + h_{km}(T_{ij})^m{}_l=0~,
\ee
which, for an invertible antisymmetric matrix $h$, is equivalent to showing
that $T_{ij}$ span a real form $C_r$ of $sp(2r, \bC)$
\footnote{The example examined above is the special case $r=1$.
Indeed, the matrices
$X_1:=\sigma_1,~X_2:=-i\sigma_2,~X_3:=\sigma_3$ generate the algebra
(\ref{euripides}) which, over the complex numbers, is isomorphic to
$C_1:=A_1$.}. If $h$ is degenerate one obtains a contraction
of ${\cal G}^s_0$.

\noindent $\bullet$ In the case where the number of Killing spinors
is odd ($N=2r+1$), $h_{ij}$
is necessarily degenerate and
without loss of generality we can take it to be of the form
\be
h=\pmatrix{0&0&\dots &0\cr 0&{}&{}&{}&\cr
~\vdots&~~~~&h_{i'j'}\cr
0&{}&{}&}~,
\ee
where $i',j'=2,\dots 2r+1$. The algebra (\ref{descartes})
becomes
\bea
{}[T_{1j},T_{mn}]&=&h_{mj}T_{1n}+h_{nj}T_{1m}\nn\\
{}[T_{i'j'},T_{m'n'}]&=&4h_{(m'|(i'}T_{j')|n')}~.
\eea
The generators
$\{T_{1m}: m=1,\dots, N\}$ span a Heisenberg algebra $H_r$ with central
generator $T_{11}$. $H_r$ is an ideal of ${\cal G}_0^s$, while
--as in the case examined above--
the generators $\{T_{i'j'}\}$ form a subalgebra of ${\cal G}_0^s$ isomorphic
to a real form $C_r$ of $sp(2r, \bC)$ or a contraction thereof. The total
algebra is the semidirect sum ${\cal G}_0^s=C_r\oplus_s H_r$.


\newsection{Supersymmetric  M-brane configurations}

Planar brane probes in Minkowski space are associated with
supersymmetry projectors. This is most easily seen
using the kappa-symmetry projector of world-volume actions. It has
been observed  in \cite{kallosh} that kappa-symmetry projectors
for all branes are associated with {\sl hermitian} matrices $\Gamma$
\be
{\rm tr}~ \Gamma=0~~~~~~~ \Gamma^2=1~.
\label{kap}
\ee
In the case of planar M-branes, the $\Gamma$'s are the elements in a Clifford
algebra associated with the world-volume forms of the branes.
The space of supersymmetry projectors has a ring structure inherited
from that of the Clifford algebra multiplication and addition.
A first test on whether a configuration of branes is supersymmetric
is to consider it as a probe in  Minkowski spacetime and
find whether the projectors associated with the branes are compatible,
i.e. whether they mutually commute and 
whether they have common eigenvalues.
If they do, then the
number of eigenspinors with eigenvalue one is the number of supersymmetries
preserved by the configuration.
Many such supersymmetric brane
configurations have been found, see e.g.~ \cite{gppkt, eric, ohtapkt, gpgg, Gaunta},
in applications in string- and M-theory,

In an M-brane configuration
let us define as `linearly independent' branes,
 those branes whose world-volumes are associated
with linearly independent elements
in the Clifford algebra. To our knowledge, it is not known
what is the maximal number
of linearly
independent branes that are allowed in a supersymmetric configuration.
We shall show that in M-theory the maximal number is thirty one and
that such a configuration
preserves one supersymmetry.
The Clifford algebra elements associated with the world-volume forms
of an M-wave,
$
MW: ~0~1~,
$
an M2-brane,
$
M2:~0~1~2~,
$
 an M5-brane,
$
M5:~0~1~2~3~4~5~,
$
and
an M-theory KK-monopole,
$
MK:~0~1~2~3~4~5~6~,
$
which we call them collectively M-branes,
 are
\be
\Gamma^{01}~,~~~ \Gamma^{012}~,~~~~\Gamma^{012345}~,~~~ \Gamma^{0123456}~,
\label{gggg}
\ee
respectively. The numbers  denote the world-volume
directions of the M-branes.
Denoting (\ref{gggg}) collectively with $\Gamma$, the Killing spinors
satisfy $\Gamma\epsilon=\epsilon$ and the supersymmetry projector is
${1\over \sqrt{2}} (1_{32}+\Gamma)$.
Observe that all the  matrices in (\ref{gggg})  are hermitian and satisfy (\ref{kap}).
Since in addition are real, they are also symmetric.
These properties imply that all
the Clifford algebra  elements (\ref{gggg}) are in $sl(32, \bR)$.
In addition,
these properties imply that each element in (\ref{gggg}) 
is non-degenerate and
has eigenvalues $\pm 1$ with equal number
of positive and negative eigenvalues.
Therefore to find the maximal number of linearly independent branes which can appear
in a supersymmetric configuration, we have to determine the  maximal number
of mutually commuting elements in $sl(32,\bR)$ which are hermitian
and satisfy (\ref{kap}). Since it is required that the projectors
of the branes of a supersymmetric configuration commute, the associated
world-volume forms of the branes  lie in a Cartan subalgebra
(CSA) of $sl(32,\bR)$. So the maximal number of linearly independent M-branes in a
supersymmetric configuration is 31.

To show that the maximal number of linearly independent
 M-branes  is precisely 31, we have to
demonstrate
that there is a basis in CSA of $sl(32,\bR)$ spanned by hermitian, traceless and
non-degenerate matrices which square to identity.
It can be arranged that all such matrices are diagonal
\be
K(\lambda_1, \dots, \lambda_{32})
=\pmatrix{\lambda_1&0&0&\dots &0\cr 0&\lambda_2&0&\dots &0\cr
~~\vdots&~~~~&\vdots\cr
0&0&0&\dots &\lambda_{32}}~,
\ee
$\lambda_\alpha=\pm 1$, $\alpha=1, \dots, 32$, and
$\sum_{\alpha=1}^{32}\lambda_\alpha=0$.
Any such matrix $K$ can be written as an element of the CSA of $gl(32, \bR)$. Indeed
$K=y^a m_{a, a+1}$,  where $m_{a, a+1}$ is the standard basis of the CSA of $gl(32, \bR)$
as in the appendix A provided that
\be
\lambda_1=  y^1~, ~~ \lambda_r=-y^{r-1}+y^r~, ~r=2, \dots, 31,
~~ \lambda_{32}=-y^{31}~.
\ee
Clearly, this equation can be solved as
\be
y^r=\sum^r_{s=1} \lambda_s~, r=1, \dots, 31~.
\ee
So $K$ is in the CSA. Conversely, any basis vector $m_{a, a+1}$ of the CSA
can be written in terms of matrices $K$. For example consider the basis
vector $m_{1,2}$. This can be written as
\be
{1\over 2}
(K(1, -1, \lambda_3, \dots, \lambda_{32})+K(1, -1, -\lambda_3, \dots, -\lambda_{32}))
\ee
and similarly for the rest. This shows that the maximal number of linearly
independent M-branes in a supersymmetric configuration is 31.

It remains to show that such configuration preserves one supersymmetry.
This can be easily seen by observing that $\lambda_1$ in all the matrices
of a basis can be chosen to be $+1$. This
is because if in one of the matrices $K$
has $\lambda_1=-1$, we can take $-K$ as a basis vector. Thus the spinor
\be
\epsilon=\pmatrix {1\cr 0\cr \vdots\cr 0}
\ee
is an eigenspinor of all the projectors with eigenvalue one. So the configuration
preserves at least one supersymmetry.
In addition there is no other eigenspinor which is linearly independent
from $\epsilon$ and has the same eigenvalue. To see this, suppose that there
is such an eigenspinor $\eta$. Without loss of generality, we can assume that
\be
\eta=\pmatrix {0\cr 1\cr 0\cr \vdots\cr 0}~.
\ee
In that case the $K$ matrices which have both $\epsilon$ and $\eta$
as eigenspinors, are of the form
\be
K(1,1, \lambda_3, \dots, \lambda_{32})~.
\ee
However such matrices do not span a basis of the 
CSA of $sl(32, \bR)$. For example,
it is not possible to construct the basis vector $m_{1,2}$.
Therefore, the configuration with 31 linearly independent branes
preserves one supersymmetry.

Given a configuration of 31 M-branes which preserves one-supersymmetry, there
are another 30 configurations of 31 branes and anti-branes
which also preserve one supersymmetry. These
configurations are made from the   branes that are involved in the original
configuration after some of them are replaced by their anti-branes.
This is because,
as we have seen, the world-volume elements of the branes of the
original configuration can be simultaneously diagonalized with eigenvalues
$\pm 1$. So any spinor $\epsilon$ with only one non-vanishing entry is
an eigenspinor of all world-volume elements with eigenvalues $\pm 1$.
The eigenvalue $+1$ is associated with the branes and the eigenvalue
$-1$ is associated with the anti-branes.

An example of a  supersymmetric M-brane configuration with 31
linearly independent  branes is
\bea
M2&:&~0~1~2
\cr
M2&:&~0~~~~~3~4
\cr
M2&:&~0~~~~~~~~~5~6
\cr
M2&:&~0~~~~~~~~~~~~~7~8
\cr
M2&:&~0~~~~~~~~~~~~~~~~~~9~10
\cr
M5&:&~0~~~2~~~4~~~6~~~8~~~10
\cr
M5&:&~0~~~2~3~~~5~~~~~8~~~10
\cr
M5&:&~0~~~2~~~4~5~~~7~~~~~10
\cr
M5&:&~0~1~~~~~4~~~6~7~~~~~10
\cr
M5&:&~0~1~~~~~4~5~~~~~8~~~10
\cr
M5&:&~0~~~2~3~~~~~6~7~~~~~10
\cr
M5&:&~0~~~2~3~~~5~~~7~~~9~
\cr
M5&:&~0~1~~~~~4~~~6~~~8~9~
\cr
M5&:&~0~1~~~3~~~~~6~~~8~~~10
\cr
M5&:&~0~~~2~~~4~5~~~~~8~9~
\cr
M5&:&~0~~~2~3~~~~~6~~~8~9~
\cr
M5&:&~0~~~2~~~4~~~6~7~~~9~
\cr
M5&:&~0~1~~~~~4~5~~~7~~~9~
\cr
M5&:&~0~1~~~3~~~~~6~7~~~9~
\cr
M5&:&~0~1~~~3~~~5~~~~~8~9~
\cr
M5&:&~0~1~~~3~~~5~~~7~~~~~10
\cr
MK&:&~0~~~~~~~~~5~6~7~8~9~10
\cr
MK&:&~0~~~~~3~4~~~~~7~8~9~10
\cr
MK&:&~0~~~~~3~4~5~6~~~~~9~10
\cr
MK&:&~0~~~~~3~4~5~6~7~8~~
\cr
MK&:&~0~1~2~~~~~~~~~7~8~9~10
\cr
MK&:&~0~1~2~~~~~5~6~~~~~9~10
\cr
MK&:&~0~1~2~~~~~5~6~7~8~~~~
\cr
MK&:&~0~1~2~3~4~~~~~~~~~9~10
\cr
MK&:&~0~1~2~3~4~~~~~7~8~~~~
\cr
MK&:&~0~1~2~3~4~5~6~~~~~~~~
\label{mot}
\eea


One can verify that all the
associated world-volume elements are mutually commuting,
hermitian and satisfy (\ref{kap}).
The above brane configuration preserves one supersymmetry. Observe
that this
 example of  supersymmetric M-brane configuration does not involve
 the M-theory pp-wave.

There is up to conjugation  a unique CSA of $sl(32, \bR)$
which is spanned by hermitian traceless matrices.
Therefore {\sl all} the supersymmetric configurations
of M-branes with 31 linearly independent branes
can be constructed by taking {\sl different bases} in the CSA of $sl(32, \bR)$  which
can be written in terms of hermitian matrices that satisfy (\ref{kap}).
In general though such bases will not be associated with simple forms
as in (\ref{mot}). It is expected that there are supersymmetric
configurations with 31 M-branes that involve the M-theory pp-wave and
the bound state of an M2-brane within an M5-brane.
In the latter case, the supersymmetry projector is a linear
combination of an M2- and an M5-brane projector \cite{gppktb}.
This classifies all supersymmetric brane
configurations with 31 linearly independent M-branes.
It also classifies all supersymmetric brane configurations with
less than 31 linearly independent M-branes. This is because
all such configurations can be constructed from those with 31
branes by removing some of the linearly independent elements
of the Clifford algebra.

The action of $SL(32, \bR)$ does {\sl  not} preserve
the space of supersymmetric M-brane configurations.
To see this consider  a supersymmetric configuration of $31$ linearly independent
 M-branes as in (\ref{mot}), and
let $\{\Gamma_I: I=1, \dots, 31\}$
be the world-volume elements of these branes in the Clifford algebra.
Next take
\be
\Gamma'_I=A\Gamma_I A^{-1}~,
\ee
where $A\in SL(32, \bR)$.
{}The matrices $\{\Gamma'_I: I+1, \dots, 31\}$ satisfy (\ref{kap}).
Requiring in addition that $\{\Gamma'_I: I+1, \dots, 31\}$ be hermitian,
we find that
\be
\Gamma_I (A^{\rm t} A)=(A^{\rm t} A)\Gamma_I~, ~~~I=1, \dots, 31~.
\label{ancon}
\ee
Since the matrix $A^{\rm t}A$ commutes with all the elements of the CSA of $sl(32, \bR)$,
$A^t A$ is in the maximal torus of $sl(32, \bR)$. Therefore
we conclude that not all elements of $SL(32, \bR)$ preserve the space
of supersymmetric planar M-brane configurations. If $A\in SO(32,\bR)\subset SL(32,\bR)$,
then (\ref{ancon}) is satisfied. So $SO(32,\bR)$ preserves  the space
of supersymmetric planar M-brane configurations but $SL(32, \bR)$ does not.
If there are solutions of eleven-dimensional
supergravity which preserve one supersymmetry some of them may have
the interpretation of  configurations above with 31 branes, see also \cite{Gaunt}.

\vskip 0.5cm
\noindent {\bf Acknowledgements}

\medskip

\noindent 
We thank C. Hull for explaining to us the results of \cite{hull}.  
DT thanks the European Union, grant number HPRN-2000-00122, which includes
the Department of Physics, Queen Mary College University of London,
as a sub-contractor.

\vskip 0.5cm

\newpage
\setcounter{section}{0}
\setcounter{subsection}{0}

\appendix{Some properties of $sl(32, \bR)$}
The Lie algebra $sl(32, \bR)$
is isomorphic to $M^0_{32}(\bR)$. A basis of $sl(32, \bR)$
can be constructed as follows:
Define $m_{a,b}$
to be the matrix which entry $1$ at the $a, b$ position and
zero otherwise, i.e.
$$
(m_{a,b}){}^c{}_d=\delta_a^c \delta_{bd}~.
$$
Then define a basis in $sl(32,\bR)$ as follows:
\bea
H_a&=&m_{a, a}-m_{a+1, a+1}~,~~~~a=1, \dots, 31
\cr
E^+_{a,b}&=& m_{a,b}~,~~~~~~~~~~~~~~~a<b
\cr
E^-_{a,b}&=& m_{a,b}~,~~~~~~~~~~~~~~~a>b
\eea
The $H$ generators span a  Cartan subalgebra of $sl(32, \bR)$, $E^+$ are
the step generators along the positive roots while $E^-$ are the step
generators along the negative roots. The step generators along the simple
roots are $E^+_a=E^+_{a, a+1}$ for $a=1, \dots, 31$.

Because of (\ref{frieza}),  all the gamma matrices can be written
in the $m_{ab}$ basis. To make this
explicit, consider the following basis of gamma matrices
\bea
\Gamma^0&=&e^0\otimes \gamma_9~, ~~ \Gamma^1=e^1\otimes \gamma_9~,
\cr
\Gamma^2&=&e^2\otimes \gamma_9~,~~
\Gamma^{i+2}=1_2\otimes \gamma^{i}~, i=1, \dots, 8~,
\eea
where $\{\gamma^i: i=1,2,\dots, 8\}$ are the gamma matrices associated
with ${\rm Cliff}(\bR^8)$,  $\gamma_9=\gamma^{12\dots 8}$
and
\bea
e^0=\pmatrix{0&-1\cr 1&0}~, ~~e^1=\pmatrix{1&0\cr 0&-1}~,
 ~~e^2=\pmatrix{0&1\cr 1&0}~.
 \eea
It is known that $Spin(8)$ admits two inequivalent real eight-dimensional
chiral spinor representations. These combine to a sixteen-dimensional
 Majorana representation.
An explicit expression for the $\gamma^i$ matrices in the latter representation is
\bea
\gamma^1&=&e^1\otimes e^0\otimes I_1~, ~~ \gamma^2=e^1\otimes e^0\otimes I_2~,~~
\gamma^2=e^1\otimes e^0\otimes I_3~,
\cr
\gamma^4&=&e^1\otimes e^0\otimes J_1~, ~~ \gamma^5=e^1\otimes e^0\otimes J_2~,~~
\gamma^6=e^1\otimes e^0\otimes J_3~,
\cr
\gamma^7&=&1_2\otimes e^1\otimes 1_4~, ~~\gamma^8=1_2\otimes e^2\otimes 1_4~,
\eea
where $I_1, I_2, I_3$ and $J_1, J_2, J_3$ are  bases  in $\Lambda^{2+}(\bR^4)$
and $\Lambda^{2-}(\bR^4)$, respectively. Explicitly, we have
\bea
(I_r)_{0s}&=&\delta_{rs}~,~~ (I_r)_{st}=\epsilon_{rst}~, ~~~~r,s,t=1,\dots,3~,
\cr
(J_r)_{0s}&=&\delta_{rs}~,~~ (J_r)_{st}=-\epsilon_{rst} ~.
\eea
It is straightforward to see that all the $\gamma^i$ matrices are real,
traceless and hermitian. All the gamma matrices $\{\Gamma^A; A=0, \dots, 10\}$
can be expressed in terms of the $m_{ab}$ basis of $sl(32, \bR)$. In particular,
we have
\be
\Gamma_A= X_A^{ab} m_{ab}~
\ee
for some numerical coefficients $X_A^{ab}$. Consequently, all the products
of gamma matrices can be written in the $m_{ab}$ basis as
\be
\Gamma_{A_1A_2\dots A_n}=X_{A_1A_2\dots A_n}^{ab} m_{ab}~,
\ee
where again $X$ are numerical coefficients.
Conversely, (\ref{frieza}) implies that we can express $m_{ab}$
in terms of skew-symmetric products
of gamma matrices. The above discussion
in particular implies that the supercovariant curvature ${\cal R}$
can be written in terms of the $m_{ab}$ basis.

\newsection{Spinorial Lie Derivatives and Killing-Yano tensors}

It is well-known that given a Killing vector field, one can define
a spinorial Lie derivative which acts as a derivation on the
space of $\Lambda^*(M)\otimes S$. Here we shall define a spinorial
Lie derivative with respect to a vector p-form, $Y$, i.e. a section of
 $T(M)\otimes \Lambda^p(M)=\Lambda_1^p(M)$.

 It is well known that given $Y\in \Lambda^p_1(M)$, we can define a derivation
 on the space of forms as
 \be
 {\cal L}_Y=i_Y d\omega+(-1)^p d i_Y \omega
 \ee
 where $\omega\in \Lambda^q(M)$, $d$ is the exterior derivative and
 \be
 i_Y\omega= {1\over p! (q-1)!} Y^b{}_{a_1\dots a_p} \omega_{b a_{p+1}\dots a_{p+q-1}}
 dx^{a_1}\wedge\dots\wedge dx^{a_{p+q-1}}~
 \ee
 In analogy with the spinorial derivative along a Killing vector, we
 take
 \be
 {\cal L}_Y\epsilon= Y^b \nabla_b \epsilon+ Q_{ab} \Gamma^{ab} \epsilon~,
 \ee
where we have suppressed the form indices of $Y$ and $Q$ is a section of
$\Lambda^p_2$ to be determined.
Demanding that
\be
{\cal L}_Y(\Gamma\epsilon)=(-1)^p \Gamma {\cal L}_Y\epsilon
\ee
we find
\be
(\nabla_a Y_{b, c_1, \dots c_p}-4 Q_{c_1\dots c_p, ab})
dx^{c_1}\wedge\dots\wedge dx^{c_p}\wedge dx^a=0~.
\label{consist}
\ee
If $Y\in \Lambda^0_1$, we can take $Q_{ab}={1\over4}\nabla_a Y_b$. Since
$Q_{ab}$ is skew-symmetric in $a,b$, consistency requires that
\be
\nabla_aY_b+\nabla_b Y_a=0
\ee
and so $Y$ is a Killing vector. If $Y\in \Lambda^1_1$, the most
general solution of (\ref{consist}) is
\be
Q_{c,ab}={1\over8} (\nabla_c Y_{a,b}-\nabla_c Y_{b,a}+
\nabla_a Y_{c,b}-\nabla_b Y_{c,a}
+\nabla_a Y_{b,c}-\nabla_b Y_{a,c})~,
\label{democritus}
\ee
where $Y_{a,b}=g_{ac} Y^c{}_b$. If $Y_{ab}=-Y_{ba}$, then
\be
Q_{c, ab}={1\over4} \nabla_c Y_{ab}~.
\ee

More generally, in order to solve (\ref{consist}), we define
\be
T_{b,c_1\dots c_{p+1}}:=\nabla_{[c_1|}Y_{b,|c_2\dots c_p]}
\ee
and expand in irreducible parts,
\be
T_{b,c_1\dots c_{p+1}} = T^{(p+1,1)}_{b,c_1\dots c_{p+1}}
+T^{(p+2,0)}_{bc_1\dots c_{p+1}}+(p+1)g_{b[c_1}T^{(p,0)}_{c_2\dots c_{p+1}]}~,
\ee
where
\bea
T^{(p,0)}_{c_1\dots c_{p}}&=&{1\over D-p}~T^f{}_{,fc_1\dots c_p}\nn\\
T^{(p+2,0)}_{c_1\dots c_{p+2}}&=&T_{[c_1,c_2\dots c_{p+2}]}\nn\\
T^{(p+1,1)}_{b,c_1\dots c_{p+1}}&=&T_{b,c_1\dots c_{p+1}}
-T^{(p+2,0)}_{bc_1\dots c_{p+1}}-(p+1)g_{b[c_1}T^{(p,0)}_{c_2\dots c_{p+1}]}~.
\label{heraclitus}
\eea
Note that an irreducible $(p,k)$-tensor
$V^{(p,k)}$ satisfies
\be
V^{(p,k)}_{[b_1\dots b_p,c_1]\dots c_k}=
g^{b_1c_1}V^{(p,k)}_{b_1\dots b_p,c_1\dots c_k}=0, ~~~~p\geq k~.
\ee
Similarly we expand,
\bea
Q_{ab,c_1\dots c_p}&=&Q^{(p,2)}_{ab,c_1\dots c_p}
+Q^{(p+2,0)}_{abc_1\dots c_p}
+p(p-1)~g_{a[c_1|}g_{b|c_2}Q^{(p-2,0)}_{c_3\dots c_p]}\nn\\
&&+p~(g_{a[c_1|}Q^{(p-1,1)}_{b,|c_2\dots c_p]}-
g_{b[c_1|}Q^{(p-1,1)}_{a,|c_2\dots c_p]})\nn\\
&&+p~(g_{a[c_1|}Q^{(p,0)}_{b|c_2\dots c_p]}-
g_{b[c_1|}Q^{(p,0)}_{a|c_2\dots c_p]})\nn\\
&&+(Q^{(p+1,1)}_{a,bc_1\dots c_p}-Q^{(p+1,1)}_{b,ac_1\dots c_p})
~.
\label{anaximandros}
\eea
Equation (\ref{consist}) is equivalent to
\bea
Q^{(p+2,0)}&=&-{1\over 4}~T^{(p+2,0)}\nn\\
Q^{(p+1,1)}&=&-{p+1\over 4p}~T^{(p+1,1)}\nn\\
Q^{(p,0)}&=&{p+1\over 4p}~T^{(p,0)}~,
\label{anaximenes}
\eea
Plugging (\ref{anaximenes}) into
(\ref{anaximandros}), we get the most general solution
to (\ref{consist}),
\bea
Q_{ab,c_1\dots c_p}&=&-{1\over 2}T_{[a,b]c_1\dots c_{p}}
+{1\over 4}T_{[c_1|,ab|c_2\dots c_{p}]}\nn\\
&&+Q^{(p,2)}_{ab,c_1\dots c_p}
+p(p-1)~g_{a[c_1|}g_{b|c_2}Q^{(p-2,0)}_{c_3\dots c_p}\nn\\
&&+p~(g_{a[c_1|}Q^{(p-1,1)}_{b,|c_2\dots c_p]}-
g_{b[c_1|}Q^{(p-1,1)}_{a,|c_2\dots c_p]})~,
\eea
where $Q^{(p,2)},~Q^{(p-2,0)},~Q^{(p-1,1)}$ are arbitrary.
Note that for $p=1$ the above expression reduces to
(\ref{democritus}).

If we set
\be
Q_{c_1\dots c_p, ab}={1\over4}\nabla_a Y_{b,c_1\dots c_p}~,
\ee
consistency  requires that
\be
\nabla_{(a} Y_{b),c_1\dots c_p}=0~,
\ee
so that $Y$ is a Killing-Yano tensor.

The anticommutator of odd generators of the  asymptotic supersymmetry algebra
of a spacetime with M2- and M5-branes is \cite{pktdd} has central terms
which
are brane charges. One may be tempted to introduce similar terms in the
symmetry superalgebra of a supersymmetric background,
\be
\{Q_i, Q_j\}=T_{ij}+ C_{ij}+G_{ij}~,
\ee
where $C$ and $G$  are the analogues of the M2-and M5-brane charges 
respectively. If $C$ and $G$
are central elements, then the commutator
of $C$ and $G$ with $Q$ would have to vanish.
These commutators may be geometrically computed
using the spinorial Lie derivatives along the two- and five-forms associated
with the Killing spinors. However a preliminary  computation has shown that such
spinorial Lie derivatives do not vanish.


\end{document}